\documentclass[12pt]{iopart}

\ifx\pdftexversion\undefined
  \usepackage[dvips]{graphicx}
\else
  \usepackage[pdftex]{graphicx}
\fi

\usepackage{hyperref}

\newcommand{\gws}{gravitational waves~}
\newcommand{\gw}{gravitational wave~}
\newcommand{\ee}[1]{\!\times\!10^{#1}}

\begin{document}

\title{Searching for \gws from known pulsars}

%\author{Matthew Pitkin, R\'ejean J. Dupuis and Graham Woan for the LIGO
%Scientific Collaboration}
\author{Matthew Pitkin for the LIGO Scientific Collaboration}
\address{Department of Physics and Astronomy, Kelvin Building,
University of Glasgow, Glasgow G12 8QQ, UK}
\ead{matthew@astro.gla.ac.uk}

\begin{abstract}
We present upper limits on the amplitude of \gws from 28 isolated pulsars
using data from the second science run of LIGO. The results are
also expressed as a constraint on the pulsars' equatorial
ellipticities. We discuss a new way of presenting such ellipticity
upper limits that takes account of the uncertainties of the pulsar
moment of inertia. We also extend our previous method to search
for known pulsars in binary systems, of which there are about 80
in the sensitive frequency range of LIGO and GEO 600.
%, that includes the system dependent binary time delays in the analysis.
\end{abstract}

\maketitle

\section{Introduction}
Over the last few years the LIGO and GEO\,600 \gw detectors have
had three (S1, S2 and S3) and two (S1 and S3) science runs
respectively. Data from S1 have been used to set upper limits on
\gw emission from a variety of possible sources \cite{S1PulPaper,
S1InspiralPaper, S1BurstPaper, S1StochPaper}. Here we concentrate
on a search for \gws from known pulsars which has been extended
both in terms of analysis method and number of sources from that
presented in \cite{S1PulPaper}.

S1 (23 August - 9 September 2002) was the {\it first} major data
taking period and a search for \gws from only one pulsar (the
fastest millisecond pulsar J1939+2134) was performed
\cite{S1PulPaper}. This search used both a time domain and a
frequency domain analysis method. For S2 (14 February - 14 April
2003) the number of pulsars being searched for increased to 28
\cite{S2PulPaper}, with the criterion that the pulsars were
isolated and had spin frequencies within the sensitive region of
the instruments, i.e. $f > 25$\,Hz or \gw frequencies of
$f_\textrm{\tiny{GW}} > 50$\,Hz, assuming emission from a triaxial
neutron star \cite{JKS}. The time domain analysis technique was
considered the optimal search for S2 where the source position and
frequency parameters were known. The analysis and results from S2,
including possible astrophysical interpretations, will be
summarised in this paper.

Analysis of the S3 run (31 October 2003 - 9 January 2004) is underway and has
the aim of extending the number of pulsars in the search to all those with
$f > 25$\,Hz ($\sim 110$) including those within binary systems. The additions
to the analysis needed to incorporate the binary systems will be discussed in
more detail in this paper. We will also discuss how the results on neutron
star ellipticity could  be interpreted as an exclusion region on a moment of
inertia - ellipticity plane.

%The main point of note is that these results have been providing the best
%{\it direct} upper limits we have on neutron star ellipticities.

\section{S2 summary}
\subsection{Analysis}
S2 data were available for analysis from all three LIGO interferometers.
Frequency and positional information for all the isolated pulsars with $f >
25$\,Hz can be found on the Australia Telescope National Facility (ATNF)
\cite{ATNF} online catalogue. However for some pulsars this information lacked
the accuracy required to be sure that the search was coherent over the period of
the science run, and additional timing over the epoch of S2 was requested and
obtained for 18\footnote{The Crab pulsar (J0534+2200) data was taken from the
online Jodrell Bank Monthly ephemeris
\url{http://www.jb.man.ac.uk/research/pulsar/crab.html}.} of these objects
\cite{KramerAndLyne}. In total this gave us 28 pulsars with reliable information
upon which to reconstruct their phase evolution. Of these 28 pulsars, 14 are in
globular clusters. The list also includes the Crab pulsar (J0534+2200) and the
fastest millisecond pulsar J1939+2134 targeted in S1.

The analysis method is described in detail in \cite{RejeansThesis}. In summary
we perform a time domain heterodyne of the data with the known phase evolution
of  the pulsar signal. After heterodyning, any signal in the data would vary
only with the beam pattern \cite{JKS} of the detector and its form would depend
on the four unknown parameters of \gw amplitude ($h_0$) and polarisation angle
($\psi$), pulsar orientation angle ($\iota$) and the signal's initial phase
($\phi_0$). The heterodyned data is then filtered and re-binned from the
detector sample rate of 16384\,Hz to $1/60$\,Hz. We then determine the
probability distribution functions of the parameter values using a Bayesian
inference technique, as shown in \cite{RejeansThesis}.

\subsection{Hardware injections}
During S2 we injected two artificial pulsar signals into the three
LIGO interferometers for 12 hours by modulating one mirror of each
via the actuation control signal. This was done to perform a
validation of the search pipeline from as far up the chain as
possible. The injections increased our confidence in the phase
calibration of the detectors, and allowed for a joint coherent
analysis of data from all the detectors.

% insert S2 injection figure

\subsection{Results and astrophysical interpretation}
The 95\% upper limits on $h_0$ %(i.e. the value of $h_0$ which, with $h_0=0$,
%bounds 95\% of the probability)
are summarised in Table.~\ref{h0hist}.
\begin{table}
\begin{center}
\begin{tabular}{c|c}
$h_0^{95\%}$ & no. of pulsars \\
\hline \hline
$1\ee{-24} < h_0 \le 5\ee{-24}$ & 20 \\
$5\ee{-24} < h_0 \le 1\ee{-23}$ & 4 \\
$h_0 > 1\ee{-23}$ & 4 \\
\hline
\end{tabular}\caption{The 95\% upper limits on $h_0$ for the 28 pulsar searched
for with the S2 run.}\label{h0hist}
\end{center}
\end{table}
We can recast this upper limit on $h_0$ as an upper limit on the pulsar's
ellipticity, $\epsilon$, calculated via
\begin{equation}\label{ellipticity}
\epsilon \simeq 0.237\frac{h_0}{10^{-24}}\frac{r}{1\,\textrm{kpc}}\frac{1\,\textrm{
Hz}^2}{f^2}\frac{10^{38}\,\textrm{kg}\,\textrm{m}^2}{I_{zz}},
\end{equation}
where $r$ is the distance to the pulsar in kpc, $f$ is its spin frequency, and
$I_{zz}$ is its principal moment of inertia \cite{S2PulPaper}. The ellipticity
limits are summarised in Table~\ref{ellipticityhist} using the canonical value
of $10^{38}\,\textrm{kg}\,\textrm{m}^2$ for the moment of inertia and assuming
no error on the pulsar's distance.
\begin{table}
\begin{center}
\begin{tabular}{c|c}
ellipticity $\epsilon$ & no. of pulsars \\
\hline \hline
$1\ee{-6} < \epsilon \le 1\ee{-5}$ & 4 \\
$1\ee{-5} < \epsilon \le 1\ee{-4}$ & 16 \\
$\epsilon > 1\ee{-4}$ & 8 \\
\hline
\end{tabular}\caption{The upper limits on $\epsilon$ for the 28 pulsar searched
for with the S2 run.}\label{ellipticityhist}
\end{center}
\end{table}
The lowest value of strain upper limits ($1.7\ee{-24}$) is associated with pulsar
J1910-5959D and of ellipticity ($4.5\ee{-6}$) with pulsar J2124-3358. The
upper limit for the Crab pulsar, at $h_0 = 4.1\ee{-23}$, is only a factor of $\sim 30$ 
greater than the canonical limit inferred from simple spin-down 
arguments\footnote{This limit is imposed by energy conservation and assumes that the 
rotational kinetic energy is only lost to \gws}. For all other pulsars the upper limits 
are a few orders of magnitude higher than their canonical limits.

Although the inferred ellipticities are well above those allowed by the
majority of conventional neutron star equations of state they are approaching
the region of astrophysical interest. Indeed, for the lowest pulsar
ellipticities ($\sim 4.5\ee{-6}$) we are starting to reach the range permitted
by at least one exotic theory of neutron star structure \cite{bumps}. It
should be stressed that these results are the first {\it direct} upper limits on
\gw emission for 26 of the 28 pulsars, with previous upper limits for pulsar
J1939+2134 and the Crab pulsar given by \cite{S1PulPaper} and \cite{CrabUL}
respectively. For the pulsars within globular clusters the spin-down can be
masked by local Doppler shifts caused by the cluster dynamics. This makes
placing a spin-down based upper limit difficult unless the pulsar motions within
the clusters can be found independently. These \gw interferometer upper limits are 
inherently independent of the cluster dynamics.

\section{S3 analysis}
We expanded the search in the S3 analysis to include all pulsars
with $f > 25$\,Hz. These include $\sim 70$ pulsars within binary (or more
complex) systems so it is important to account for the extra Doppler and
relativistic time delays due to the pulsar's orbital motion. An
advantage over S2 is the availability of GEO\,600 data in addition to LIGO data.

\subsection{Binary pulsar signal}
In the case of an isolated pulsar the signal received at the detector needs to
be corrected to the solar system barycentre (SSB) by calculating the Doppler
delays and other relativistic effects. This is possible as the pulsar's position
is known and we have very good solar system ephemerides. Of course it is simply
the reverse procedure to that used to determine the positional and frequency
parameters of the pulsar from radio data, so the reconstruction is generally of
good quality. The motion of a pulsar in a binary system adds a number of Doppler
and relativistic time delays:
\begin{equation}
\Delta{}T_\textrm{\tiny{bin}} = \Delta_\textrm{\tiny{R}} +
\Delta_\textrm{\tiny{E}} +
\Delta_\textrm{\tiny{S}} + \Delta_\textrm{\tiny{A}},
\end{equation}
where $\Delta_\textrm{\tiny{R}}$ is the Roemer delay (light travel time),
$\Delta_\textrm{\tiny{E}}$ is the Einstein delay due to special relativistic
effects, $\Delta_\textrm{\tiny{S}}$ is the general relativistic Shapiro delay
and $\Delta_\textrm{\tiny{A}}$ is the abberation delay caused by the pulsar's
rotation. These delay changes can be far more pronounced than those from the
Earth's motion, with up to a 0.03 Hz frequency shift. The time delays are
parameterised by properties of the binary system including its period,
eccentricity, angular velocity, time of periastron, projected semi-major axis
and several relativistic parameters depending on the nature of the system. These
parameters are found by fitting the radio observations (using the standard TEMPO
data reduction package \cite{TEMPO}) to various binary models (e.g.
\cite{TaylorAndWeisberg}). The model to which the binary system is
fitted will depend on how relativistic it is or which parameters you wish to
extract. The 70 binary system pulsars fall mainly into two models: 32 in the low
eccentricity (ELL1 model \cite{Lange:2001}) and 33 in the Blandford-Teukolsky
(BT model \cite{TaylorAndWeisberg}). Four fall into the highly relativistic
Damour-Deruelle (DD \cite{TaylorAndWeisberg}) model and a further one into
the Blanford-Teukolsky-2-Planet (BT2P) model, although this can be adequately
fit using the simpler BT model. With this additional binary information provided
the full set of pulsars can be included in the search. So far new timing
information for the majority of these pulsars has been provided
\cite{KramerAndLyne} and the analysis is underway.

\subsection{Hardware injections}
We injected ten artificial isolated pulsar signals, with a wide range of
signal parameters into the LIGO interferometers. One other signal was injected
into both GEO\,600 and LIGO. The signal strengths ranged from marginally
detectable to very strong. These signals have been successfully extracted using
the search pipeline \cite{RejeansThesis}. For the strongest signals, agreement 
between the injected and recovered parameter values has been to within a few percent.

% maybe include fig from Rejean's thesis.

\section{Moment of inertia - ellipticity plane}
\begin{figure}[!htbp]
\begin{center}
\includegraphics[width=0.5\textwidth]{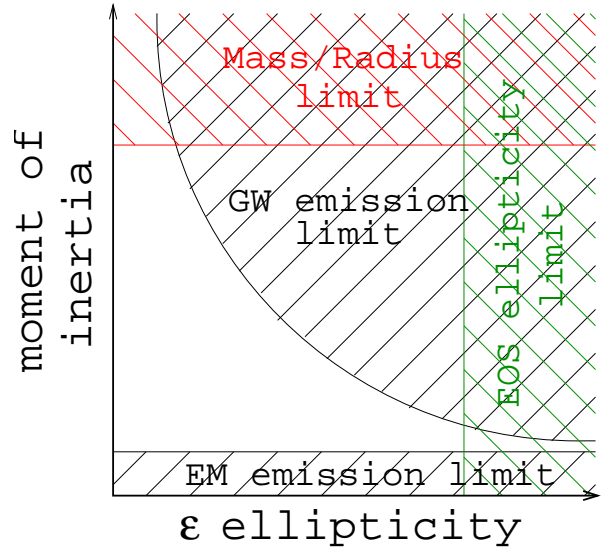}
\caption{This figure shows a rough representation of the regions in the
moment of inertia $I_{zz}$ - ellipticity $\epsilon$ plane that can be excluded
via various methods. The electromagnetic emission of a pulsar can set a lower
limit on the moment of inertia by equating the EM emission with the rotational
energy loss of the pulsar \cite{BejgerAndHaensel:2003}. The various equations of
state for neutron stars can constrain the mass/radius relation and therefore
moment of inertia. Equations of state will also put limits on the maximum 
allowable ellipticity of the neutron star. A limit can be set from upper limits
on gravitational wave emission.}
\label{Ieplane}
\end{center}
\end{figure}

So far we have used the canonical value for the neutron star moment of inertia
($10^{38}\,\textrm{kg}\,\textrm{m}^2$) when calculating the pulsar
ellipticities. This is the moment of inertia of a $1.4\,\textrm{M}_\odot$ sphere
of uniform density and radius 10\,km. The true value can vary by factors of a
few under different models for the neutron star equation of state.
Equation~\ref{ellipticity} shows that $h_0$ can be used to set an upper limit on
the neutron star's quadrupole moment, $I_{zz}\epsilon$, that is independent of
the actual value of $I_{zz}$. This value can then be used to form an exclusion
region in the $I_{zz} - \epsilon$ plane (see Fig.~\ref{Ieplane}).
This plane can provide exclusion regions on both the moment of inertia and
ellipticity or can be used to read off an upper limit on ellipticity for a
preferred value of $I_{zz}$ such as the canonical value.

\begin{figure}[!htbp]
\begin{center}
\includegraphics[width=0.5\textwidth]{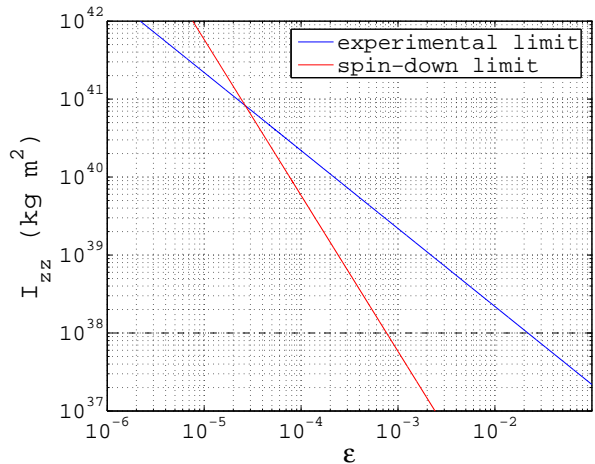}
\caption{The areas above the full lines are the excluded regions on the
moment of inertia - ellipticity plane for the Crab pulsar as obtained from the
S2 $h_0$ upper limit and the spin-down arguments. These assume a distance to the
Crab pulsar of 2 kpc.}
\label{CrabIEplane}
\end{center}
\end{figure}

Figure~\ref{CrabIEplane} shows such an exclusion region for the Crab pulsar
upper limit from S2.
An exclusion region can also be set using the spin-down based upper limit
argument, at
\begin{equation}
I_{zz}\epsilon = \frac{5|2\pi\dot{f}|c^5}{32G(2\pi{}f)^5}\frac{1}{\epsilon}.
\end{equation}
The point at which the experimental results and the spin-down result cross shows
the the point at which the experimental results beats that of spin-down. For the
S2 Crab pulsar result it can be seen that the experimental result only beats the
spin-down result for unfeasably large values of $I_{zz}$ ($> 8\ee{40}
\textrm{kg}\,\textrm{m}^2$). The upper limit on the ellipticity for the
experimental result and assuming $I_{zz} = 10^{38}\,\textrm{kg}\,\textrm{m}^2$
is $\sim 2\ee{-2}$.

\section{Future work}
The analysis of S3 data is presently underway targeting most known
pulsars within the band. In addition we will be incorporating
other sources of error into the current Bayesian framework,
including systematic calibration errors in the value of $h_0$ and
pulsar distance errors into the value of $\epsilon$ or the $I_{zz}
- \epsilon$ plane exclusion region. The S3 results promise to give
a reasonable improvement over S2 for some pulsars. In particular,
the Crab pulsar result should only be a factor of a few above the
spin-down limit.

The next science run of LIGO and GEO\,600 (S4) is due to be underway in early
2005 providing more data to analyse. The pipeline used to obtain the results in
this paper is at a mature stage and will be applied directly to the S4 data.

% references

\end{document}